\documentclass[twocolumn,preprintnumbers,amsmath,amssymb,aps,nofootinbib,floatfix,superscriptaddress,showpacs]{revtex4}
\usepackage{graphicx} %-> wrapper fuer includegraphics grfguide.ps
\usepackage{color}

\newcommand{\vev}[1]{\langle{#1}\rangle}
\newcommand{\qq}{\vev{\bar q q}}

\graphicspath{ 
{./} 
{./../figures/pqm_model_figures/}
}

\def\di{\displaystyle}

\def\bg{\begin{eqnarray}\begin{array}{rcl}\displaystyle}
\def\eg{\end{array} &\di    &\di   \end{eqnarray}}
\def\bm#1{\begin{eqnarray}\begin{array}{#1}\di}
\def\bmo#1{\begin{eqnarray*}\begin{array}{#1}\di}
\def\bml#1#2{\begin{eqnarray}\begin{array}{#1}\label{#2}\di}
\def\bgo{\begin{eqnarray*}\begin{array}{rcl}\displaystyle}
\def\ego{\end{array} &\di    &\di \nonumber  \end{eqnarray*}}

\def\btensor#1#2{\renew\left#1\begin{array}{#2}\di}
\def\brtensor#1#2#3{\ren#3\left#1\begin{array}{#2}}
\def\botensor#1#2{\renew\left#1\begin{array}{#2}}
\def\etensor#1{\end{array}\right#1}

\def\eq#1{(\ref{#1})}
\def\Eq#1{Eq.~(\ref{#1})}

\def\tr{{\rm tr}}
\def\Tr{{\rm Tr}}

\def\s0#1#2{\mbox{\small{$ \frac{#1}{#2} $}}}
\def\0#1#2{\frac{#1}{#2}}

\def\dr{{D\!\llap{/}}\,}

\def\CP{{\mathcal P}}

\def\CU{{\mathcal U}}

\def\ren#1{\renewcommand{\arraystretch}{#1}}

\def\renew{\renewcommand{\arraystretch}{1}}

\begin{document}
\title{The Phase Structure of the Polyakov--Quark-Meson Model}
\author{B.-J. Schaefer}
\email{bernd-jochen.schaefer@uni-graz.at}
\affiliation{Institut f\"ur Physik, Karl-Franzens-Universit\"at Graz,
  Universit\"atsplatz 5, A-8010 Graz, Austria} 
\author{J.M.~Pawlowski}
\email{j.pawlowski@thphys.uni-heidelberg.de}
\affiliation{Institut f\"ur Theoretische Physik, Universit\"at
  Heidelberg, Philosophenweg 16, D-69120 Heidelberg, Germany}
\author{J. Wambach} 
\email{wambach@physik.tu-darmstadt.de}
\affiliation{Institut
  f\"ur Kernphysik, TU Darmstadt, Schlo{\ss}gartenstr.~9, D-64289 Darmstadt, Germany}
\affiliation{Gesellschaft
  f\"ur Schwerionenforschung mbH, Planckstr.~1,  D-64291 Darmstadt, Germany}

%General properties of QCD (dynamics, confinement, etc.)

\begin{abstract}
  The relation between the deconfinement and chiral phase transition
  is explored in the framework of an Polyakov-loop-extended two-flavor
  quark-meson (PQM) model. In this model the Polyakov loop dynamics is
  represented by a background temporal gauge field which also couples
  to the quarks. As a novelty an explicit quark chemical potential and
  $N_f$-dependence in the Polyakov loop potential is proposed by using
  renormalization group arguments. The behavior of the Polyakov loop
  as well as the chiral condensate as function of temperature and
  quark chemical potential is obtained by minimizing the grand
  canonical thermodynamic potential of the system. The effect of the
  Polyakov loop dynamics on the chiral phase diagram and on several
  thermodynamic bulk quantities is presented.
\end{abstract}

\maketitle 
 
\newpage 

%%%%%%%%%%%%%%%%%%%%%%%%%%%%%%%%%%%%%%%%%%%%%%%%%%%%%%%%%%%%%%%%%%%%%%%%%%%%%
% Introduction
%%%%%%%%%%%%%%%%%%%%%%%%%%%%%%%%%%%%%%%%%%%%%%%%%%%%%%%%%%%%%%%%%%%%%%%%%%%%%
\section{Introduction}

Driven by the heavy-ion programs at GSI, CERN SPS, RHIC and soon the
LHC there is strong interest in the properties of strongly interacting
matter at extreme temperatures and baryon densities. Ultimately these
have to be understood on the basis of Quantum Chromodynamics (QCD),
which governs the strong interaction sector of the Standard Model.
  
QCD at zero temperature and density is well-established by now both
numerically and analytically. Besides numerical evaluations on
discrete space-time lattices,
e.g.~\cite{Karsch:2001cy,Greensite:2003bk}, functional methods based
on the Functional Renormalization Group (FRG),
e.g.~\cite{FRG,Schaefer:2006sr}, and Dyson-Schwinger equations (DSE),
e.g.~\cite{DSE}, have been used to elucidate our understanding of the
theory of strong interactions. If applicable, lattice computations
give high numerical accuracy with small truncation errors. Functional
methods have their advantages if it comes to the deep infrared, the
simple explanation of physical mechanisms as well as the inclusion of
chiral dynamical quarks. The last years have seen a very fruitful
interaction between the different methods leading to a largely
quantitative understanding of QCD at vanishing temperature and density
even though the full understanding of the confinement mechanism and
its relation to spontaneous chiral symmetry breaking is not yet
settled.

At finite temperature and in particular at finite baryon density or
chemical potential $\mu_B$ the situation is much less clear. At finite
$\mu_B$ lattice computations have principal limitations due to the
complex action which hampers stringent theoretical evaluations of the
QCD phase diagram, in particular the possible critical endpoint of the
line of first-order transitions from first principles. Functional
methods have been used to obtain results for pure Yang-Mills at finite
temperature, as well as the hadronic sector of QCD at finite
temperature and density, see e.g.~\cite{FRG,Schaefer:2006sr}. A full
QCD study is hampered by the fact that the gauge sector, i.e.~the
confinement-deconfinement phase transition in pure Yang-Mills, is not
fully resolved yet: the potential for the order parameter, the
Polyakov loop, does not lead to a phase transition in perturbative
computations, e.g.~\cite{Weiss:1980rj,Engelhardt:1997pi,Gies:2000dw},
but recently this gap has been closed within a non-perturbative flow
study \cite{BGP}. Hence, a first principle approach to QCD at finite
temperature and density with functional methods is in reach. In our
opinion this opens a way to continuing the fruitful interaction
between the different methods that has already proven so successful at
vanishing temperature. 
 
A first step towards a full QCD study with functional methods is done
by studying effective Lagrangian models which are constructed from the
non-perturbative Yang-Mills effective potential and effective hadronic
models. In recent years a promising realization of this idea has been
put forward, based on lattice results for the thermodynamic potential
in pure Yang-Mills (YM) theory and universality
arguments~\cite{Pisarski:2000eq,Pisarski:2006hz,fukushima:2000,
  Fukushima:2003fm,%
  Fukushima:2003fw,Fukushima:2002bk,Meisinger:2002kg,Meisinger:2003uf,%
  Ratti:2005jh, Ratti:2004ra,Sasaki:2006ww, Meisinger:1995ih}. Pure YM
theory corresponds to the heavy-quark limit of QCD in which the
Polyakov loop expectation value serves as an order parameter for
confinement. This approach results in an effective scalar
$Z_{N_c}$-theory whose physical minima are the vacuum expectation
values of the Polyakov loop. It has been observed that a thermodynamic
potential for the Polyakov loop can be constructed where the
parameters are fitted to precise finite-temperature lattice data for
the equation of state (EoS) in the heavy-quark limit. This very
successful theory simulates the calculated first-order
confinement-deconfinement phase transition at finite temperature. For
two light flavors, on the other hand, QCD exhibits an (almost) exact
chiral symmetry and is believed to be in same universality class as
the $O(4)$ model~\cite{Pisarski:1983ms}. An effective realization of
this symmetry is provided by the Nambu-Jona-Lasinio (NJL) model or,
upon bosonization, the quark-meson (QM) model. It is therefore natural
to combine both aspects of QCD by coupling light chiral quarks to the
Polyakov loop field. This results in the PNJL
model~\cite{Meisinger:1995ih,Mocsy:2003qw,%
  Fukushima:2003fw,Meisinger:2001cq,Ratti:2005jh} or the PQM model,
which has the benefit of renormalizability, and a simpler linkage to
full QCD, see e.g.\ \cite{Gies:2002hq}. Calculations for the
thermodynamic potential and the resulting phase structure will be
performed at the mean-field level as was done in similar analyses
using the PNJL model. Eventually, however, we plan to include
fluctuations using RG-techniques~\cite{Pawlowski2006}.

Even at the mean-field level there remain some open issues. First of
all, the Polyakov loop potentials suggested so far, are fixed at
vanishing $\mu_B$. In this case, the expectation value $\Phi$ of the
Polyakov loop operator and that of its adjoint, $\bar\Phi$ are linked
by complex conjugation. At finite chemical potential this relation is
lost and the effective potential depends on the two independent
variables $\Phi$ and $\bar\Phi$. Extensions to finite $\mu_B$ have
therefore to be carefully evaluated. Moreover, the flavor and density
dependence of the pure Yang-Mills potential has not been explored as
yet. We will show that it is possible to extract the qualitative
behavior in the fully coupled system by perturbative arguments, as
well as physical consistency arguments.

The outline of the paper is as follows: in the next section we
introduce the Polyakov loop variable and discuss its effective
potential derived from lattice data. The Polyakov loop potential is
then coupled to the quark-meson model which defines the
Polyakov--quark-meson model. The grand canonical thermodynamic
potential of this model is derived in mean-field approximation, and
the choice of model parameters is discussed.
Sec.~\ref{sec:results_thermo} is devoted to thermodynamical
applications, in particular we evaluate the pressure, quark number
density and quark number susceptibility. Moreover, the phase
structure, i.e. the chiral and confinement-deconfinement phase
transition is explored. Subsequently, the influence of the Polyakov
loop potential on the thermodynamics is investigated and in
Sec.~\ref{sec:conclusion} concluding remarks are drawn.

%%%%%%%%%%%%%%%%%%%%%%%%%%%%%%%%%%%%%%%%%%%%%%%%%%%%%%%%%%%%%%%%%%%%%%%%%%%%%
\section{Polyakov--quark-meson model}
\subsection{Polyakov loop potential}

A key observable in QCD at finite temperature is the Polyakov loop.
Its expectation value serves as an order parameter for confinement in
the heavy-quark limit. The Polyakov loop operator is a Wilson loop in
the temporal direction and reads
\begin{eqnarray}\label{eq:pol} 
  \CP(\vec x)=\CP \exp \left( i \int_0^\beta d\tau A_0(\vec x , \tau)
  \right)\,,  
\end{eqnarray} 
where $\CP$ stands for path-ordering and $A_0 (\vec x , \tau)$ is the
temporal component of the Euclidean gauge field
$A_\mu$~\cite{Polyakov:1978vu,Susskind:1979up,Borgs:1983yk,Kuti:1980gh}.
The color trace of \eq{eq:pol} in the fundamental representation
$\tr_c \CP(\vec x)$ is the creation operator of a static quark at
spatial position $\vec x$.  Periodic boundary conditions ensure gauge
invariance of \Eq{eq:Phi} up to center elements. This goes hand in
hand with the fact that the temporal or Weyl gauge $A_0=0$ cannot be
achieved for periodic boundary conditions. Its physical interpretation
is best seen in Polyakov gauge, where the temporal component of the
gauge field is time-independent, $A_0(\vec x,\tau)=A_0^c(\vec x)$,
and is in the Cartan sub-algebra (see e.g.~\cite{Reinhardt:1997rm,
  Ford:1998bt, Ford:1998mq, Jahn:1998nw}). Hence, within this gauge
the Polyakov loop operator simplifies to
\begin{eqnarray}\label{eq:simple} 
\CP(\vec x)= \exp \left( i\beta A^c_0 (\vec x) \right) \,, 
\end{eqnarray} 
with $A^c_0 (\vec x)=A_0^{(3)}(\vec x)\tau_3+ A_0^{(8)}(\vec
x)\tau_8$. This results in a simple relation between the Polyakov loop
and the temporal component of the gauge field,
\begin{eqnarray}\label{eq:A_0fromCP}
A^c_0(\vec x)=-i\left(\partial_\beta \CP(\vec x)\right) 
\CP^\dagger(\vec x)\,. 
\end{eqnarray} 

The normalized Polyakov loop variable $\Phi(\vec x)$ and its hermitian
(charge) conjugate $\bar \Phi(\vec x)$ are defined as the thermal
expectation value of the color trace of the Polyakov loop operator
\eq{eq:pol}
\begin{eqnarray}\label{eq:Phi}
  \Phi(\vec x)=\0{1}{N_c}\langle \tr_c \CP(\vec x)\rangle_\beta \,,\quad
  \bar\Phi(\vec x)=\0{1}{N_c}\langle \tr_c \CP^\dagger(\vec x)
  \rangle_\beta\,.  
\end{eqnarray} 
We emphasize again that the traces are taken in the fundamental
representation. $\Phi(\vec x)$, $\bar \Phi(\vec x)$ are complex scalar
fields. Their mean values, i.e.~the solution of the quantum equations
of motion, are related to the free energy of a static, infinitely
heavy test quark (antiquark) at spatial position $\vec x$. The order
parameter $\Phi(\vec x)$ vanishes in the confined phase where 
the free energy of a single heavy quark diverges. In the deconfined
phase it takes a finite value. 
%%%%%%%%%%%%%%%%%%%%%%%%%%%%%%%%%%%%%%%%%%%%%%%%%%%%%%%%%%%%%%%%%%%%%%%%%%%%%
The correlation function of two Polyakov loop variables is related to
the free energy $F_{q \bar q}$ of two color sources $q$ and $\bar q$
with spatial separation $\vec r = \vec x - \vec y$ as
\begin{equation}\label{eq:cor}
  \frac{ 1}{N_c^2}\vev{ \tr _c \CP(\vec x) \tr _c 
    \CP^\dagger(\vec y)} _\beta =
  e^{-\beta F_{q \bar q } (\vec r)}\,. 
\end{equation} 
The dependence on $\vec r$ allows one to extract the string tension.
The cluster decomposition property (locality) enforces that for
infinite distance the correlation between a quark and anti-quark
vanishes and we arrive at
\begin{eqnarray}
\frac{ 1}{N_c^2}\vev{ \tr _c \CP(\vec x) \tr _c 
    \CP^\dagger(\vec y)} _\beta\to \Phi(\vec x) \bar \Phi(\vec y)\,.
\end{eqnarray} 
These properties provide the Polyakov criterion of confinement at
finite temperature. It is linked to the center $Z_{N_c}$ symmetry of
the $SU(N_c)$ gauge group: a gauge transformation that is periodic up
to a center element, leads to
\begin{eqnarray} \label{eq:Z3} 
\Phi\to z\Phi,\quad z \in Z_{N_c}\,  .
\end{eqnarray} 
Thus, the confining phase is center symmetric, whereas in the
deconfined phase center symmetry is spontaneously broken.

In summary, the confinement-deconfinement phase transition is
characterized by the mean value $\Phi=0$ in the confined phase and a
finite non-zero value in the deconfined phase.
%%%%%%%%%%%%%%%%%%%%%%%%%%%%%%%%%%%%%%%%%%%%%%%%%%%%%%%%%%%%%%%%%%%%%%%%%%%%%
In the presence of dynamical quarks, the free energy of a
quark-antiquark pair does not diverge anymore, and the order parameter
is always non-vanishing. For finite quark chemical potential the free
energies of quarks and antiquarks are different. Since $\Phi$ is
related to the free energy of quarks and the hermitian (charge)
conjugate $\bar \Phi $ to that of antiquarks, their modulus in general
differs, i.e.~$\bar \Phi\neq\Phi^\dagger$. In pure Yang-Mills theory
the mean values $\Phi $, $\bar \Phi $ are given by the minima of the
effective Polyakov loop potential $\CU (\Phi,\bar\Phi)$. It can be
constructed from lattice data for the expectation values $\Phi $,
$\bar
\Phi$~\cite{Pisarski:2000eq}.
Here we use a polynomial expansion
in $\Phi$, $\bar\Phi$ up to quartic terms. This leads to an effective
potential $\CU $ in terms of the moduli $| \Phi |$ and $|\bar\Phi |$,
the product $\Phi\bar\Phi$, and in $\Phi^3$, $\bar\Phi^3$ related to
the $Z_3$ symmetry. The $U(1)$-symmetric part of $\CU$ is a
Ginzburg-Landau type potential.

In pure Yang-Mills we can restrict ourselves to fields with the same
modulus, $| \Phi |=|\bar\Phi |$. With this additional constraint we
have $\Phi\bar\Phi=|\bar \Phi |^2$ and we can drop one of the
$U(1)$-invariants in the expansion. This has been used in
Ref.~\cite{Pisarski:2000eq} where the potential is only expanded in
$\Phi\bar\Phi$, $\Phi^3$ and $\bar\Phi^3$. Alternatively one can use
the moduli and drop the $\Phi\bar\Phi$-term. We conclude that the
general effective potential in this approximation reads as
\begin{eqnarray}\nonumber 
&&\hspace{-1.4cm} \0{\CU(\Phi,\bar\Phi)}{T^4}=-\0{b_2}{4}
\left(|\Phi|^2+|\bar\Phi|^2 \right)
\\ 
&&\hspace{+.3cm}
-\0{b_3}{6}(\Phi^3+\bar\Phi^3)+\0{b_4}{16}
\left(|\Phi|^2+|\bar\Phi|^2\right)^2\,.\label{eq:polpot} 
\end{eqnarray} 
The expansion coefficients are fixed to reproduce thermodynamic
lattice results for the pure YM sector as in
Refs.~\cite{depietri-2005-LAT2005,Heinzl:2005xv,deForcrand:2001nd,%
Wozar:2006fi, Ratti:2005jh,Ratti:2004ra}. This leads to
temperature-independent coefficients $b_3 = 0.75$ and $b_4 = 7.5$, and
a temperature-dependent one $b_2$ with
\begin{eqnarray}\label{eq:b2}
\hspace{-4ex} 
  b_2(T) &=& a_0  + a_1 \left(\frac{T_0}{T}\right) + a_2
  \left(\frac{T_0}{T}\right)^2 + a_3 \left(\frac{T_0}{T}\right)^3\, 
\end{eqnarray} 
where $a_0 = 6.75$, $a_1 = -1.95$, $a_2 = 2.625$, $a_3 = -7.44$, $b_3
= 0.75$ and $b_4 = 7.5$. The effective potential \eq{eq:polpot} can be
augmented by logarithmic terms, see e.g.~Ref.~\cite{Ratti:2007jf}.
This will be discussed in Sec.~\ref{sec:finitemu}.

The potential \eq{eq:polpot} with the above parameters has a
first-order phase transition at the critical temperature $T_0 = 270$
MeV.

%%%%%%%%%%%%%%%%%%%%%%%%%%%%%%%%%%%%%%%%%%%%%%%%%%%%%%%%%%%%%%%%%%%%%%%%%%%%%
\subsection{Coupling to the quark-meson sector}
\label{sec:meanfield}

The hadronic properties of low-energy QCD with light flavors are
effectively incorporated by a chiral quark-meson model. Here the local
$SU(N_c)$ gauge invariance of the underlying QCD is replaced by a
global symmetry in the original quark-meson model which results in the
loss of the confinement property. The QM model shows a chiral phase
transition at realistic temperatures~e.g.~\cite{Schaefer:2006sr,
  quark_meson}. In the limit of massless quarks the order parameter of
the chiral phase transition is the quark condensate $\qq$. For
realistic up- and down quark masses chiral symmetry is broken
spontaneously and also explicitly in the vacuum resulting in a finite
chiral condensate $\qq$. Due to the lack of confinement in this model
single quark states are already excited at low temperatures in the
chirally broken phase, see e.g.~\cite{Schaefer1999} resulting in an
unrealistic EoS near the phase transition. Since the constituent quark
masses are much larger than that of the pion the meson dynamics
dominates at low temperatures and the predictions from chiral
perturbation theory are reproduced.
 
By combining the Polyakov loop model with the QM model chiral as well
as confining properties of QCD are included. This promising approach
has been put forward in \cite{ Fukushima:2003fm,Fukushima:2003fw,
  Fukushima:2002bk, Meisinger:2002kg, Ratti:2005jh,Megias:2004,
  Megias:2005qf, Megias:2002vr, Megias:2006bn} and significantly
improved the EoS near the phase boundary. The integration over the
gluonic degrees of freedom in the presence of a homogeneous background
for the temporal component $A_0$ yields the Polyakov loop potential
and the mesonic terms of the chiral QM model. Thus, the dynamical
quark sector of QCD is included by integrating out the quarks in the
presence of mean background fields. This finally leads to a coupled
Polyakov--quark-meson model with an interaction potential between
quarks, mesons and the Polyakov loop variables $\Phi$, $\bar\Phi$. To
leading loop order this potential is provided by the Dirac determinant
in the presence of the mean fields.

The generalized Lagrangian of the linear QM model for $N_f=2$ light
quarks $q=(u,d)$ and $N_c=3$ color degrees of freedom coupled to a
spatially constant temporal background gauge field reads
\begin{eqnarray} \label{eq:qmmodel}
  {\cal L} &=& \bar{q} \,(i\dr - g (\sigma + i \gamma_5
  \vec \tau \vec \pi ))\,q 
  +\frac 1 2 (\partial_\mu \sigma)^2+ \frac{ 1}{2}
  (\partial_\mu \vec \pi)^2
  \nonumber \\
  && \qquad - U(\sigma, \vec \pi )  -{\cal U}(\Phi,\bar\Phi)\ ,   
\end{eqnarray}
where the purely mesonic potential is defined as
\begin{eqnarray} \label{eq:pot}
U(\sigma, \vec \pi ) &=& \frac \lambda 4 (\sigma^2+\vec \pi^2 -v^2)^2
-c\sigma\ \,.
\end{eqnarray}

The isoscalar-scalar $\sigma$ field and the three
isovector-pseudoscalar pion fields $\vec \pi$ together form a chiral
vector field $\vec \phi$. Without the explicit symmetry breaking term
$c$ in the mesonic potential the Lagrangian is invariant under global
chiral $SU(2)_L\times SU(2)_R$ rotations. The covariant Dirac operator
$D_\mu = \partial_\mu -i A_\mu$ in \eq{eq:qmmodel} reads
\begin{eqnarray}\label{eq:dirac}
 \dr(\Phi)=\gamma_\mu \partial_\mu-i\gamma_0 A_0(\Phi)\,.
\end{eqnarray} 
The spatial components of the gauge fields have vanishing background
i.e.~$A_\mu = \delta_{\mu 0} A_0$.

%%%%%%%%%%%%%%%%%%%%%%%%%%%%%%%%%%%%%%%%%%%%%%%%%%%%%%%%%%%%%%%%%%%%%%%%%%%%%
\subsection{Polyakov loop potential parameters}

In the presence of dynamical quarks, the running coupling $\alpha$ is
changed due to fermionic contributions. In our approximation to the
Polyakov loop potential this only leads to a modification of the
expansion coefficient $b_2$, \Eq{eq:b2}. The size of this effect
can be estimated within perturbation theory, see
e.g.~\cite{Banks:1981nn, Miransky:1996pd, Appelquist:1996dq,
  Braun:2006jd, Braun:2005uj}. At zero temperature it leads to an
$N_f$-dependent decrease of $\Lambda_{\rm QCD}$, which translates into
an $N_f$-dependent decrease of the critical temperature $T_0$ at
finite temperature. The two-loop $\beta$-function of QCD with massless
quarks is given by
\begin{eqnarray}
\label{eq:beta}
\beta(\alpha)=-b \alpha^2-c \alpha^3\,,
\end{eqnarray}
with the coefficients
\begin{eqnarray}\label{eq:coeffs}
\hspace{-4ex}
  b&=& \0{1}{6 \pi} (11 N_c-2 N_f)\,, \\
\hspace{-4ex}
c&=&\0{1}{24\pi^2}
  \left(34 N_c^2-10 
    N_c N_f -3 \0{N_c^2-1}{N_c} N_f\right)\,. 
\label{eq:coeffs2}\end{eqnarray}
Here, we have assumed a RG scheme that minimizes (part of) the
higher-order effects. This is an appropriate scheme for our mean-field
analysis. At leading order the corresponding gauge coupling is given
by 
\begin{eqnarray}
  \label{eq:coupling} 
  \alpha(p)=\0{\alpha_0}{1+\alpha_0 b \ln
    (p/\Lambda)}+O(\alpha_0^2) \,,
\end{eqnarray}
with $\alpha_0=\alpha(\Lambda)$ at some UV-scale $\Lambda$, and
$\Lambda_{\rm QCD}=\Lambda \exp(-1/(\alpha_0 b))$. At
$p=\Lambda_{\rm QCD}$ the coupling \eq{eq:coupling} exhibits a Landau
pole. At finite temperature the relation \eq{eq:coupling} allows us to
determine the $N_f$-dependence of the critical temperature $T_0(N_f)$.
For $N_f=0$ it is given by $T_0=270$ MeV which corresponds to fixing
the coupling $\alpha_0$ at the $\tau$-scale $T_\tau =1.770$ GeV and a
running coupling of $\alpha_0=0.304$ accordingly. If one keeps the
coupling $\alpha_0$ at $T_\tau$ fixed, this identification yields the
relation
\begin{eqnarray}\label{eq:relation} 
T_0(N_f)=T_\tau e^{ -1/(\alpha_0 b)}\,,
\end{eqnarray}
and Table~\ref{tab:critt} for the $N_f$-dependent critical temperature
$T_0$ in the Polyakov loop potential for massless flavors:
\begin{table}[h!]
  \begin{tabular}{c||@{\hspace{2mm}}c@{\hspace{2mm}}|@{\hspace{2mm}}
      c@{\hspace{2mm}}|@{\hspace{2mm}}c@{\hspace{2mm}}|@{\hspace{2mm}}
      c@{\hspace{2mm}}|@{\hspace{2mm}}c@{\hspace{2mm}}}
    $N_f$ & $0$ & $1$ & $2$ & $2+1$ & $3$  \\
    \hline
    \hline
    $T_0$ [MeV] & 270 & 240 & 208 & 187 & 178 \\
    \hline
  \end{tabular}
  \caption{\label{tab:critt} Critical Polyakov loop temperature $T_0$ for
    $N_f$ massless flavors.}
\end{table}

Massive flavors lead to suppression factors of the order
$T_0^2/(T_0^2+m^2)$ in the $\beta$-function. For $2+1$ flavors and a
current strange quark mass $m_s\approx 150$ MeV we obtain
$T_0(2+1)=187$ MeV. We remark that the estimates for $T_0(N_f)$ have
an uncertainty at least of the order $\pm 30$ MeV. This uncertainty
comes from the perturbative one-loop nature of the estimate and the
poor accounting for the temperature effects.  For example, with the
two loop coefficient \eq{eq:coeffs2} and concentrating on $N_f=2$ as
studied in the present work we are led to $T_0(2)=192$ MeV. 
Fortunately, the results only show a mild $T_0$ dependence.

Finally, we argue that there are no double counting effects due to the
inclusion of the Dirac determinant in the PQM and the independent
adjustment of the Polyakov loop model parameters: the Polyakov loop
potential parameters, in particular $b_2$, \Eq{eq:b2}, genuinely
depend on the running coupling, which is changed in the presence of
quarks. This effect is modeled by changing $T_0\to T_0(N_f)$ as
defined in \Eq{eq:relation}. The direct contributions to the Polyakov
loop potential which originate from the fermionic determinant
$\Omega_{\bar q q}$, \Eq{eq:Omegaintqq}, are not governed by this
redefinition, and have to be added separately.

%%%%%%%%%%%%%%%%%%%%%%%%%%%%%%%%%%%%%%%%%%%%%%%%%%%%%%%%%%%%%%%%%%%%%%%%%%%%%
\subsection{Non-vanishing chemical potential} 
\label{sec:finitemu}

A further intricacy concerns the Polyakov loop potential at finite
chemical potential \cite{Roessner:2006xn, fukushima-2006}. Then the
constraint $\bar\Phi= \Phi^\dagger$ ceases to be valid, and the
extension of \Eq{eq:polpot} to finite $\mu$ is not unique anymore. For
further details see e.g. Refs.~\cite{Dumitru:2002cf, Dumitru:2003hp}.
The leading $\mu$-dependence of the full potential stems from the
Dirac determinant, and we assume that the $\Phi,\bar\Phi$-symmetric
form of the potential \eq{eq:polpot} persists at finite $\mu$. Then
the only additional $\mu$ dependence originates from a possible $\mu$
dependence of the model parameters. This approximation certainly is
valid for small chemical potential where the $\mu$ dependence is
rather small. The remaining ambiguity concerns possible  
$\Phi\bar\Phi$-terms, that can be incorporated
into the potential \eq{eq:polpot} by the replacement
\begin{eqnarray}\label{eq:YMwithmu}
\s012 (|\Phi|^2+|\bar\Phi|^2)\to 
 \s012 \theta (|\Phi|^2+|\bar\Phi|^2)+(1-\theta)
\Phi\bar\Phi\,. \
\end{eqnarray}   
\Eq{eq:YMwithmu} leaves the potential unchanged for $\mu=0$, that is
$\bar\Phi= \Phi^\dagger$. For positive $\theta$ the potential has
unstable directions, e.g.~for vanishing $\Phi$ or $\bar \Phi$, and
large $\bar\Phi$, $\Phi$ respectively. Hence the choice $\theta=0$
possibly leads to negative susceptibilities. In
Refs.~\cite{Ratti:2005jh,Sasaki:2006ww} this choice has been used, and
the computed susceptibilities are not positive anymore
\cite{Sasaki:2006ww}. This problem has been cured in
Refs.~\cite{Sasaki:2006ww, Roessner:2006xn,Ratti:2007jf} by augmenting
the Polyakov loop potential with logarithmic terms. Effectively, this
amounts to changing the model parameters in the polynomial ansatz used
in these works. For $\theta\neq 0$ these logarithmic terms are not
necessary, due to lack of unstable directions. Furthermore, a weak
total $\mu$ dependence as well as the validity of the mean-field
analysis would hint at the preferred choice $\theta=1$. However, for
this choice the expectation value of ($\bar\Phi-\Phi$) has the wrong
sign, even though other observables show a mild $\theta$-dependence.
Clearly, this structure is related to the present mean-field
approximation. It should be possible to overcome this parameter
dependence in a fully non-perturbative setting. Here, we shall show
results for the choice $\theta=0$.

In a final step we implement a $\mu$-dependent running coupling in the
$b_2$ coefficient, analogous to the $N_f$-dependence discussed above.
Indeed, one can argue that this is a minimal necessary generalization:
without a $\mu$-dependent $b_2$ the confinement-deconfinement
phase-transition has a higher critical temperature than the chiral
phase transition at vanishing chemical potential. This is an
unphysical scenario because QCD with dynamical massless quarks in the
chirally restored phase cannot be confining since the string breaking
scale would be zero.

As for the $N_f$-dependence we resort to perturbative estimates. To
begin with we simply allow for an additional $\mu$-dependent term in
the one-loop coefficient $b$,
\begin{eqnarray}\label{eq:bmu}
  b(\mu)=\0{1}{6 \pi} (11 N_c-2 N_f)- b_\mu
\frac{\mu^2}{T^2_\tau}\,.
\end{eqnarray} 

This specific simple choice of the $\mu$-dependent part can be motivated 
by using HDL/HTL results on the effective charge~\cite{LeBellac1996}
\begin{eqnarray}\label{eq:alphasmu}
\alpha(p,T,\mu)=\frac{\alpha(p)}{1+m^2_D/p^2}\,, 
\end{eqnarray} 
with the perturbative Debye mass
$m_D^2=(N_c/3+N_f/6) g^2 T^2+N_f/(2 \pi^2) g^2 \mu^2$. The
$\mu$-derivative of the modified coupling, $\mu
\partial_\mu \alpha = b_\mu \mu^2/p^2$, can be related to a momentum
derivative $p\partial_p \alpha = -b(p,\mu) \alpha^2$. Within the
present simple approach based on a $\mu$-dependence only valid in the
perturbative regime we estimate the momentum-dependent coefficient
$b(p,\mu)$ by $b(\mu)=b(\gamma\, T_\tau,\mu)$ at an (average) momentum
scale $\gamma\, T_\tau$ with $\gamma\leq 1$.

The coefficient $b_\mu$ can be fixed such that the chiral transition
temperature and the confinement-deconfinement transition agree at some
non-vanishing $\mu$. Interestingly, it turns out that then the
transition temperatures agree for all $\mu$'s. The related value of
$b_\mu$ is provided by $\gamma\simeq 1/4$ and
\begin{eqnarray}\label{eq:bmuvalue}
  b_\mu \simeq \frac{16}{\pi} N_f \,.
\end{eqnarray} 
Inserting the $\mu$-dependent coefficient $b(\mu)$ into
~\Eq{eq:relation} then leads to an additional $\mu$-dependent $T_0$,
\begin{eqnarray}\label{eq:relationmu} 
  T_0(\mu, N_f)=T_\tau e^{ -1/(\alpha_0 b(\mu))}\,.
\end{eqnarray}
\Eq{eq:relationmu} with \eq{eq:bmuvalue} should be viewed as a rough
estimate of the $\mu$-dependence of $T_0$. We emphasise again that
this simple estimate leads to coinciding phase boundary lines for the
chiral and confinement-deconfinement transition, see
Sec.~\ref{sec:results_thermo}. For more quantitative results the
non-perturbative running of the coupling in the presence of finite
temperature and quark density has to be considered. This can be
incorporated in a self-consistent RG-setting. Moreover, one has to
resolve the uncertainties, discussed at the beginning of this section,
concerning the form of the effective potential at finite $\mu$.

Here we will present a comparison of the phase diagram with and
without $\mu$-dependent $T_0$ in Fig.~\ref{fig:phase_diagram_theta0}.
For the other results the additional $\mu$-dependence is taken into
account.

%%%%%%%%%%%%%%%%%%%%%%%%%%%%%%%%%%%%%%%%%%%%%%%%%%%%%%%%%%%%%%%%%%%%%%%%%%%%%
\section{Applications}
\label{sec:results_thermo}

The PQM model is defined by the Lagrangian~\eq{eq:qmmodel} with the
Polyakov loop potential~\Eq{eq:polpot}. The dependence of the
coefficient $b_2$~\Eq{eq:b2} on the $N_f$-dependent or ($N_f,
\mu$)-dependent running coupling $\alpha$ is governed
by~\Eq{eq:relation} and~\eq{eq:relationmu} respectively. This defines
the starting point for an investigation of the phase structure and
bulk thermodynamics of the PQM model. The thermodynamics is
characterized by the grand canonical potential which is analyzed in
mean-field approximation.

\subsection{Grand canonical potential}

The grand canonical potential in a spatially uniform system is
determined as the logarithm of the partition function, which in our
case is a path-integral over the meson and quark/antiquark fields
including the Polyakov loop. We confine ourselves to the
$SU(2)_f$-symmetric case and set $\mu \equiv \mu_u =\mu_d$. This is a
good approximation to the realistic case since flavor mixing in the
vector channel is small. Integrating over the fermions by using the
Nambu-Gor'kov formalism and introducing averaged meson fields yields
the grand canonical potential
\begin{eqnarray} \label{eq:totalpot} \Omega &=& \CU(\Phi,\bar\Phi) +
  U(\sigma) + \Omega_{\bar qq}(\Phi,\bar\Phi,\sigma)
\end{eqnarray}
with the quark/antiquark contribution
\begin{eqnarray} \label{eq:Omegaintqq}
\Omega_{\bar qq} &=& -2N_f T
\!\!\int\!\!\frac{d^3p}{(2\pi)^3} \tr _c \left\{\ln (1 + 
  {\cal P} 
  e^{-(E_p -\mu)/T}) +\qquad\ \right. \nonumber \\
&& \hspace{2.5cm} \left.\ln (1 + {\cal P^\dag}
  e^{-(E_p +\mu)/T})\right\}
\end{eqnarray}
and the purely mesonic potential
\begin{eqnarray} \label{eq:mesonpot}
U(\sigma) &=& \frac \lambda 4 (\sigma^2 -v^2)^2
-c\sigma\ .
\end{eqnarray} 

The divergent vacuum part in the quark/anti-quark contribution is
absorbed in the renormalization which is done in the vacuum. The
quark/antiquark single-quasiparticle energy is given by
\begin{eqnarray}
  E_p &=& \sqrt{\vec p^2 + m_q^2}
\end{eqnarray}
with the constituent quark mass $m_q = g \sigma$. The remaining color
trace in the quark/antiquark contribution \eq{eq:Omegaintqq} is
evaluated by using the identity $\Tr \ln A= \ln\det A$ and yields
\begin{eqnarray} \label{eq:Omegaqq}
\Omega_{\bar qq} = -2N_f T
\!\!\int\!\!\frac{d^3p}{(2\pi)^3}\hspace{3cm}&&  \\
\left\{\ln\!\! \left[1\! +\! 3 (\Phi + \bar
    \Phi e^{-(E_p-\mu)/T})e^{-(E_p-\mu)/T}\!\!+\!\!e^{-3(E_p-\mu)/T}\right]
  +\right.&&\nonumber \\
 \left.  \ln\!\! \left[1\! +\! 3 (\bar \Phi + \Phi
    e^{-(E_p+\mu)/T})e^{-(E_p+\mu)/T}\!\!+\!\!e^{-3(E_p+\mu)/T}\right]
\right\}.&&\nonumber 
\end{eqnarray}
Note, that no ultraviolet cutoff is necessary because the PQM model is
renormalizable in contrast to the PNJL model (see
e.g.~\cite{Ratti:2004ra, Ratti:2005jh}).

The equations of motion are obtained by minimizing the thermodynamic
potential~\eq{eq:totalpot} w.r.t. the three constant mean fields
$\sigma$, $\Phi$ and $\bar \Phi$:
\begin{equation}\label{eq:eom}
\left.\frac{\partial \Omega}{\partial \sigma}=\frac{\partial
  \Omega}{\partial \Phi}=\frac{\partial \Omega}{\partial
  \bar\Phi}\right|_{\sigma=\vev {\sigma} ,\Phi=\vev{\Phi},\bar{\Phi}=\vev
{\bar{\Phi}}} = 0\,.
\end{equation}
The solutions of these coupled equations determine the behavior of the
chiral order parameter $\vev \sigma$ and the Polyakov loop expectation
values $\vev \Phi$ and$\vev{\bar\Phi}$ as a function of $T$ and $\mu$.

%%%%%%%%%%%%%%%%%%%%%%%%%%%%%%%%%%%%%%%%%%%%%%%%%%%%%%%%%%%%%%%%%%%%%%%%%%%%%
\subsection{Quark-meson parameters}

The four parameters of the QM model, i.e.~$g$, $\lambda$, $v$ and
$c$, are chosen such that chiral symmetry is spontaneously broken in
the vacuum and the $\sigma$-field develops a finite expectation value
$\langle\sigma\rangle \equiv f_\pi$, where $f_\pi=93$ MeV is set to
the pion decay constant. Due to the pseudoscalar character of the
pions the corresponding expectation values vanish, $\langle {\vec
  \pi}\rangle =0$.

The Yukawa coupling constant $g$ is fixed by the constituent quark
mass in the vacuum $g= m_q/f_\pi$. Using the partially conserved axial
vector current (PCAC) relation the explicit symmetry breaking
parameter $c$ is determined by $c=m_\pi^2 f_{\pi}$, where $m_{\pi}$ is
the pion mass. The quartic coupling constant $\lambda$ is given by the
sigma mass $m_{\sigma}$ via the relation $\lambda= (m_\sigma^2
-m_\pi^2)/ ({2f_\pi^2})$. Finally, the parameter $v^2$ is found by
minimizing the potential in radial direction, yielding $v^2 = {\vev
  \sigma}^2-c/(\lambda \vev \sigma)$. For the ground state where $\vev
\sigma= f_{\pi}$ this expression can be rewritten as $v^2 = f_{\pi}^2
-m^2_\pi/\lambda$ . It is positive in the Nambu-Goldstone phase.

In the vacuum we fix the model parameters to $m_{\pi} =138$ MeV,
$m_{\sigma} =600$ MeV, $f_{\pi} =93$ MeV and $m_q = 300$ MeV which
result in $c \sim 1.77\cdot 10^6$ MeV$^3$, $v\sim 87.6$ MeV,
$\lambda \sim 19.7$ and $g \sim 3.2$.

%%%%%%%%%%%%%%%%%%%%%%%%%%%%%%%%%%%%%%%%%%%%%%%%%%%%%%%%%%%%%%%%%%%%%%%%%%%%%
\subsection{Phase structure}

The phase structure of the PQM model is determined by the behavior of
the order parameters $\sigma$, $\Phi$ and $\bar \Phi$ and of the grand
canonical potential as a function of temperature and quark chemical
potential. All numerical results have been obtained for $N_f=2$. Then
$T_0= 208$ MeV in agreement with Tab.~\ref{tab:critt}. This value is
different from that taken in Ref.~\cite{Ratti:2005jh, Sasaki:2007fi}
where $T_0 = 270$ MeV, the value of $N_f=0$. In these works $T_0 =210$
MeV has been fixed in order to compare with lattice results. The
$N_f$-dependence suggested in the present work offers a qualitative
explanation for this choice.

\begin{figure}[htb]
  \centerline{\hbox{
      \includegraphics[width=0.7\columnwidth,angle=-90]{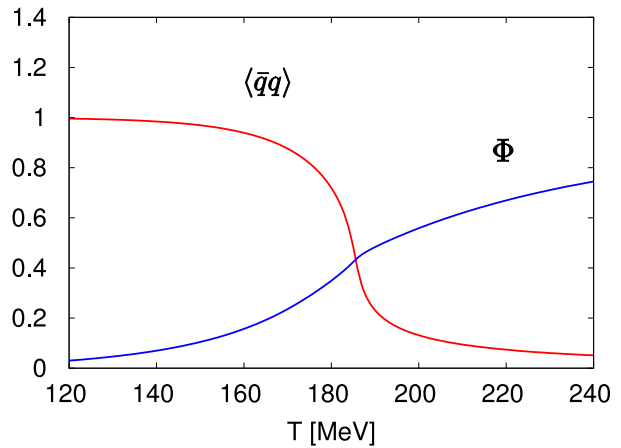}
    }}
\caption{\label{fig:condensateT0210} The normalized chiral
    condensate $\vev {\bar q q}$ and the Polyakov loop $\Phi$ as a
    function of temperature for $\mu=0$. A chiral crossover is
    found at $T\sim 180$ MeV and a deconfinement crossover at a
    similar temperature.}
\end{figure}

In Fig.~\ref{fig:condensateT0210} the temperature dependence of the
chiral condensate $\vev {\bar{q}q}$ and the Polyakov loop expectation
value $\Phi$, see \Eq{eq:Phi}, at $\mu=0$ is shown in relative units.
\vspace{.3cm}
%%%%%%%%%%%%%%%%%%%%%%%%%%%%%%%%%%%%%%%
%%%%%%%%%%%%%%%%%%%%%%%%%%%%%%%%%%%%%%%%

\begin{figure}[!htb]
  \centerline{\hbox{
      \includegraphics[width=0.7\columnwidth,angle=-90]{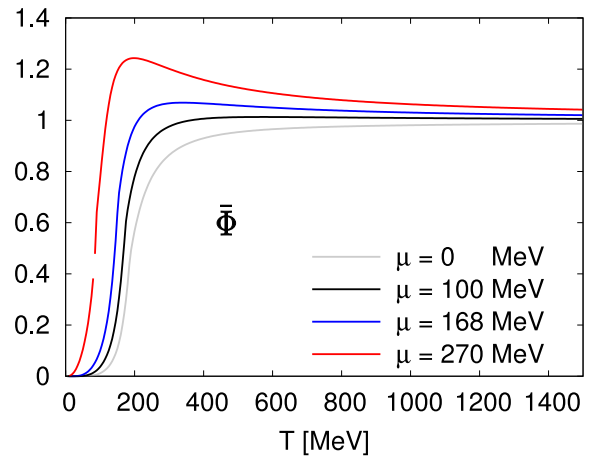}
    }}
  \caption{\label{fig:vevTinfinity} The normalized Polyakov loop
    variable $\bar \Phi$ for large temperatures for several chemical
    potentials $\mu$.}
\end{figure}

For vanishing chemical potential we have $\bar \Phi = \Phi$ as already
discussed. For $T\to \infty$ we find $\Phi \simeq 1.11$. Since the
properly normalized expectation value $\bar \Phi$ tends towards unity
we have normalized the mean fields accordingly. For temperatures at
about 200-300 MeV the normalized $\bar \Phi$ has a $\mu$-dependent
maximum and decreases for larger temperatures towards one, see
Fig.~\ref{fig:vevTinfinity}. This is in qualitative agreement with
perturbation theory which predicts an increasing $\bar \Phi$ within an
expansion around vanishing gauge fields, e.g.~\cite{Kaczmarek:2005uv,
  kaczmarek-2005-71}.

\begin{figure}[!htb]
  \centerline{\hbox{
      \includegraphics[width=0.7\columnwidth,angle=-90]{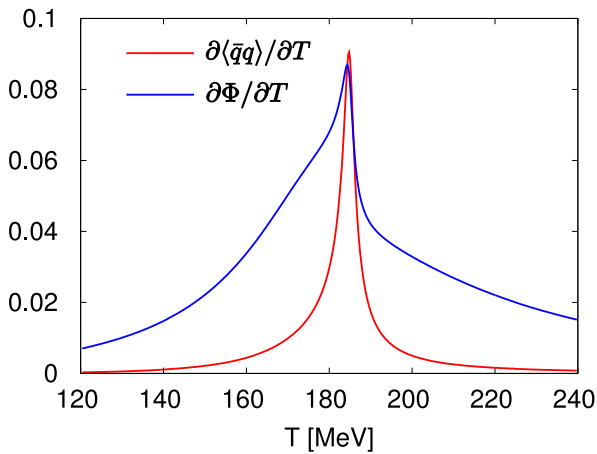}
    }}
  \caption{\label{fig:peak210} The temperature dependence of
    $\partial \vev{\bar q q}/\partial T$ and
    $\partial \Phi/\partial T$ for $\mu=0$. The Polyakov variable is
    scaled by a factor of 5.}
\end{figure}

At $\mu=0$ we find a chiral crossover temperature $T_c = 184$ MeV with
an error of $\sim \pm 14$ MeV originating in the error estimate
$\pm 30$ MeV for $T_0$. For example, using the two-loop running of the
coupling \eq{eq:coeffs2}, and hence $T_0(N_f)=192$ MeV we are led to
$T_c \sim 177$ MeV. In the presence of dynamical quarks the Polyakov
loop shows also a crossover at the same pseudo-critical temperature.
This can be read off from the peak position of
$\partial \vev{\bar q q}/\partial T$ and $\partial \Phi/\partial T$.
In Fig.~\ref{fig:peak210} these quantities are shown as function of
the temperature.

In two-flavor lattice simulations extrapolated to the chiral limit a
pseudo-critical temperature $T_c =173 \pm 8$ MeV is found using
improved staggered fermions \cite{Karsch:2001cy}. Recently, a
recalculation of the transition temperature with staggered fermions
for two light and one heavier quark mass close to their physical
values yields a $T_c = 192 \pm 7$ MeV using the Sommer parameter $r_0$
for the continuum extrapolation \cite{cheng-2006-74}. This result has
to be contrasted with another recent lattice analysis with staggered
fermions but using four different sets of lattice sizes $N_\tau = 4,6,8$
and $10$ to perform the continuum extrapolation \cite{aoki-2006-643}.
From the same physical observable this group finds a critical
temperature $T_c = 151 \pm 3$ MeV. Functional RG studies yield a
critical value of $T_c = 172^{+40}_{-34}$ MeV \cite{Braun:2006jd,
  Braun:2005uj}, where the error originates in an estimate of the
uncertainty similar to the considerations put forward here.  On the
other hand, using the same parameters for the quark-meson model
without the Polyakov loop modifications a crossover temperature of
$T_c \sim 150$ MeV emerges~\cite{Schaefer2006a}. This situation calls
for refined studies both on the lattice as well as within functional
methods to resolve the apparent quantitative inaccuracies.

\begin{figure}[!htb] 
  \centerline{\hbox{
      \includegraphics[width=0.7\columnwidth,angle=-90]{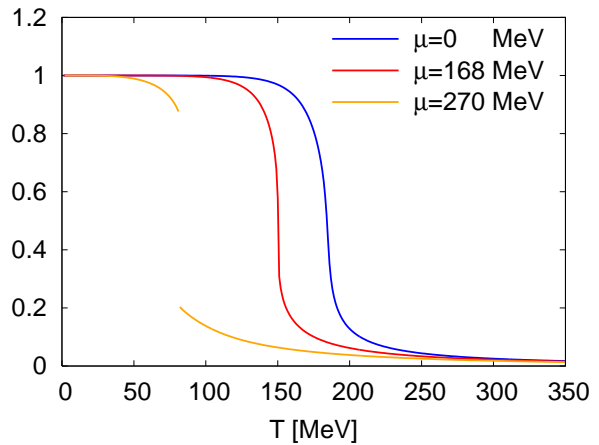}
    }}
  \caption{\label{fig:T0210} The normalized chiral quark condensate
    $\vev \sigma$ as a function of temperature for three different
    chemical potentials $\mu=0, 168, 270$ MeV. For $\mu=270$ MeV a
    first-order transition is found at $T_c \sim 81$ MeV.}
\end{figure}

\begin{figure}[!htb] 
  \centerline{\hbox{
      \includegraphics[width=0.7\columnwidth,angle=-90]{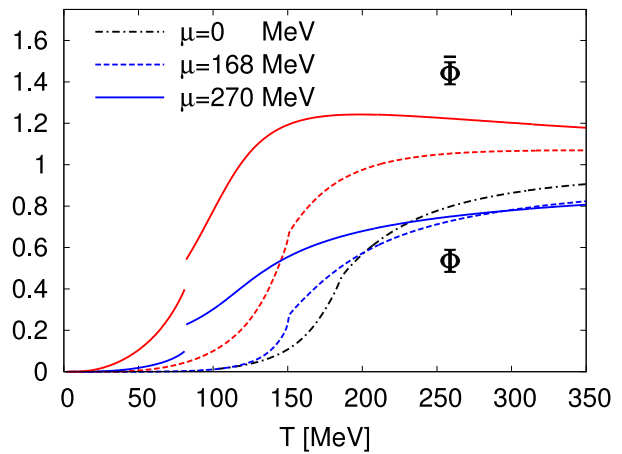}
    }}
  \caption{\label{fig:T0210phi} Same as Fig.~\ref{fig:T0210} for the
    normalized Polyakov loops $\bar \Phi$ and $\Phi$.}
\end{figure}

For finite $\mu$ the degeneracy of $\Phi$ and $\bar \Phi$ disappears.
The corresponding order parameters as function of temperature for
several chemical potentials are collected in Figs.~\ref{fig:T0210} and
\ref{fig:T0210phi}.
%%%%%%%%%%%%%%%%%%%%%%%%%%%%%%%%%%%%%%%%%%%%%%%%%%%%%%%%%%%%%%%%%%%%%%%%%%%%%
%%%%%%%%%%%%%%%%%%%%%%%%%%%%%%%%%%%%%%%%%%%%%%%%%%%%%%%%%%%%%%%%%%%%%%%%%%%%%
%%%%%%%%%%%%%%%%%%%%%%%%%%%%%%%%%%%%%%%%%%%%%%%%%%%%%%%%%%%%%%%%%%%%%%%%%%%%%
For finite $\mu$ the Polyakov loop $\bar \Phi$ is always larger than
$\Phi$. It has a positive slope
$\partial \bar \Phi/{\partial \mu} > 0$ for all temperatures, and
peaks at some high temperature, see Fig.~\ref{fig:vevTinfinity}. Both,
$\Phi$ and $\bar \Phi$ tend towards one for $T\to \infty$.

\begin{figure}[!htb]
  \centerline{\hbox{
      \includegraphics[width=0.7\columnwidth,angle=-90]{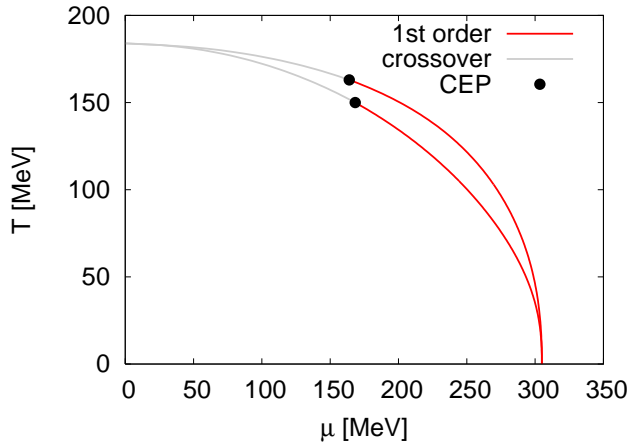}
    }}
  \caption{\label{fig:phase_diagram_theta0} Chiral phase diagrams for
    the PQM model. Upper lines for a $\mu$-independent Polyakov loop
    potential and lower lines with $\mu$-dependent corrections. The
    CEP's are approximately located at ($T_c, \mu_c$) = (163, 164) MeV
    (upper case) and at (150, 168) MeV (lower case).}
\end{figure}

Above a critical chemical potential $\mu_c = 168$ MeV all order
parameters jump at the same temperature which signals a first-order
phase transition. The critical end point (CEP) is found at
$(T_c,\mu_c) = (150,168)$ MeV. The corresponding chiral phase diagram
obtained for a $\mu$-independent $T_0 (N_f)$, \eq{eq:relation}, is
shown in Fig.~\ref{fig:phase_diagram_theta0} (upper lines).
%%%%%%%%%%%%%%%%%%%%%%%%%%%%%%%%%%%%%%%%%%%%%%%%%%%%%%%%%%%%%%%%%%%%%%%%%%%%%
%%%%%%%%%%%%%%%%%%%%%%%%%%%%%%%%%%%%%%%%%%%%%%%%%%%%%%%%%%%%%%%%%%%%%%%%%%%%%
At the critical point the chiral first-order transition line
terminates and the transition becomes second-order, which induces a
divergent quark number susceptibility. Lattice simulations are not
conclusive concerning the existence and location of the critical point
\cite{Karsch:2001cy, owe, Allton:2002zi}.

There are indications from lattice simulations at finite chemical
potential that deconfinement and chiral symmetry restoration appear
along the same critical line in the phase diagram. For the PQM model
and $\mu$-independent $T_0 (N_f)$ the coincidence of deconfinement and
chiral transition at $\mu=0$ disappears for finite $\mu$. The
deconfinement temperature is larger than the corresponding chiral
transition temperature. This is an unphysical scenario because the
deconfinement temperature should be smaller or equal to the chiral
transition temperature. When resorting to the $\mu$-dependent $T_0(\mu,
N_f)$,~\eq{eq:relationmu}, we find coinciding transition lines for the
entire phase diagram within an accuracy of $\pm 5$ MeV. For this case
the unique transition line lies slightly below the chiral one for the
$\mu$-independent choice $T_0( N_f)$. This is shown in
Fig.~\ref{fig:phase_diagram_theta0}.

%%%%%%%%%%%%%%%%%%%%%%%%%%%%%%%%%%%%%%%%%%%%%%%%%%%%%%%%%%%%%%%%%%%%%%%%%%%%%
\subsection{Thermodynamic observables}

In order to investigate the influence of the Polyakov loop on the
equilibrium thermodynamics we calculate several thermodynamic
observables. All information of the system is contained in the grand
canonical potential which is given by $\Omega$ in~\eq{eq:totalpot}
evaluated at the mean-field level.

\begin{figure}[!htb]
  \centerline{\hbox{
      \includegraphics[width=0.7\columnwidth,angle=-90]{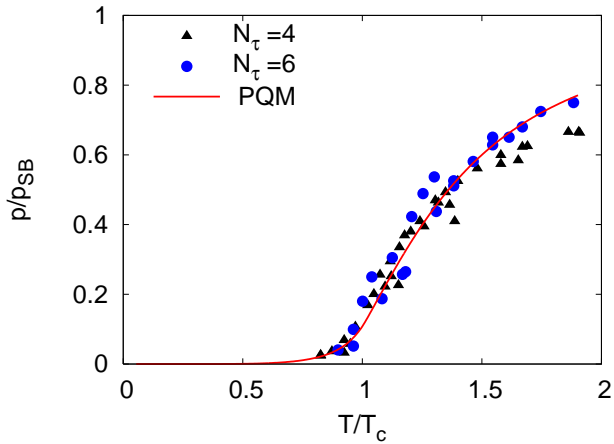}
}}
\caption{\label{fig:p_w_lattice} Scaled pressure $p/p_\mathrm{SB}$ for
  $\mu =0$. The PQM model prediction (solid line) is compared to
  lattice results for $N_\tau = 4$ and $N_\tau = 6$. Lattice data
  taken from Ref.~\cite{AliKhan:2001ek}.}
\end{figure}

We begin our analysis with the pressure of the system $p$. It is
defined as the negative of the grand canonical potential and is
normalized to vanish at $T=\mu=0$.
%%%%%%%%%%%%%%%%%%%%%%%%%%%%%%%%%%%%%%%%%%%%%%%%%%%%%%%%%%%%%%%%%%
\begin{figure}[!htb]
  \centerline{\hbox{
      \includegraphics[width=0.7\columnwidth,angle=-90]{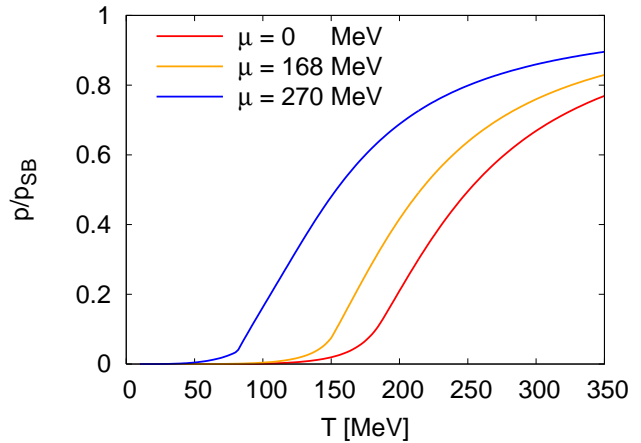}
}}
  \caption{\label{fig:pressure} Scaled pressure $p/p_\mathrm{SB}$ for
    three different quark chemical potentials, $\mu =0 , 168, 270$
    MeV. $T_c(\mu=0) = 184$ MeV.}
\end{figure}
%%%%%%%%%%%%%%%%%%%%%%%%%%%%%%%%%%%%%%%%%%%%%%%%%%%%%%%%%%%%%%%%%%%
In Fig.~\ref{fig:pressure} the pressure divided by the QCD
Stefan-Boltzmann (SB) limit is shown as function of the temperature
for three different quark chemical potentials. The values of the
chemical potentials are chosen such that one curve runs through the
critical end point (CEP) ($\mu_c = 168$ MeV) and another curve through
a first-order phase transition ($\mu = 270$ MeV). The QCD pressure in
the SB limit for $N_f$ massless quarks and $(N_c^2-1)$ massless
gluons, relevant for the deconfined phase, is given by
\begin{equation}
  \frac{ p_{\mathrm{SB}}}{T^4} = (N_c^2-1)\frac{ \pi^2}{45}
  + N_c N_f\!\!
  \left[ \frac{ 7\pi^2}{180} \!+\! \frac{ 1}{6} \left( 
      \frac{ \mu}{T}
    \right)^2 \!+\! \frac{ 1}{12\pi^2} \left( \frac{ \mu}{T} 
    \right)^4
  \right]\ .  
\end{equation}
where the first term denotes the gluonic contribution and the rest
involves the fermions. The pressure is suppressed in the confined
phase and starts to rise when deconfinement sets in. For all $T$ and
$\mu$ the pressure $p/T^4$ stays below the QCD SB limit, a feature
that is also observed in lattice calculations and other
non-perturbative approaches. For vanishing chemical potential the
pressure is a smooth function of the temperature consistent with a
crossover transition. At temperatures of twice the critical
temperature the pressure reaches approximately $80\%$ of the SB limit.
On the lattice two classes of data for the pressure obtained with a
temporal extent $N_\tau=4$ and $N_\tau=6$ at $\mu=0$ are currently
available both of which are not extrapolated to the
continuum~\cite{AliKhan:2001ek, Allton2005gk}. Our results are in
agreement with lattice simulations with a temporal extent of $N_\tau=6$
which is also closer to the continuum limit. This behavior is
demonstrated in Fig.~\ref{fig:p_w_lattice}.
%%%%%%%%%%%%%%%%%%%%%%%%%%%%%%%%%%%%%%%%%%%%%%%%%%%%%%%%%%%%%%%%%%
%%%%%%%%%%%%%%%%%%%%%%%%%%%%%%%%%%%%%%%%%%%%%%%%%%%%%%%%%%%%%%%%%%%

An increase of the chemical potential leads to an increase of the
pressure as more quark degrees of freedom are active. For a certain
chemical potential the crossover transition changes to a first-order
phase transition. In this case the pressure has a kink at the
transition point but still is a continuous function. The kink at
$T\sim 100$ MeV for the $\mu=270$ MeV curve is clearly visible
Fig.~\ref{fig:pressure}.

\begin{figure}[!htb]
  \centerline{\hbox{
      \includegraphics[width=0.7\columnwidth,angle=-90]{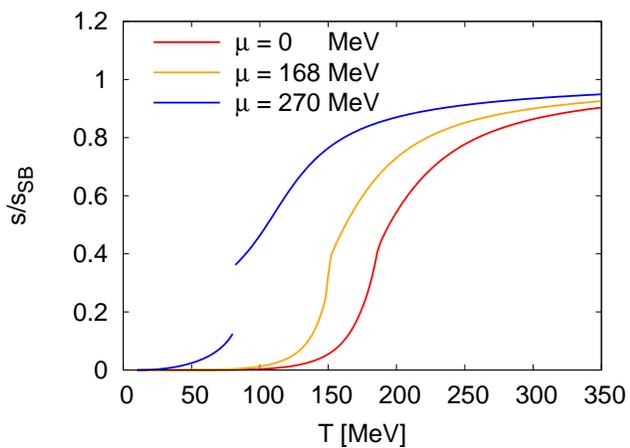}
    }}
  \caption{\label{fig:entropy} Same as described in the legend to
    Fig.~\ref{fig:pressure} for the scaled entropy $s/s_\mathrm{SB}$.}
\end{figure}

At a first-order phase transition a finite latent heat builds up. This
results in a jump of the entropy density $s$, which is defined as the
negative derivative of the grand canonical potential with respect to
the temperature. It is identical to the temperature derivative of the
pressure. In Fig.~\ref{fig:entropy} we show $s$ divided by the
corresponding QCD SB limit for the same chemical potentials as in the
preceding figure.
The low- and high-temperature behavior of this quantity can be
understood in a similar fashion as those of the pressure. It is
continuous in the vicinity of the crossover transition and reaches
less than $40\%$ of the SB limit around these temperatures. For
chemical potential values larger than the critical one a finite latent
heat emerges which further increases with the chemical potential.

Another quantity that is accessible in lattice QCD at finite chemical
potential is the pressure difference $\Delta p$. It is defined as
$\Delta p(T,\mu) = p(T,\mu) - p(T,\mu =0)$ and is Taylor-expanded
around $\mu=0$ in powers of the dimensionless quantity $\mu/T$ on the
lattice. Because odd derivatives of the free energy with respect to
$\mu$ vanish only even powers appear in this expansion. In our model
we have computed the pressure difference without referring to an
expansion.
%%%%%%%%%%%%%%%%%%%%%%%%%%%%%%%%%%%%%%%%%%%%%%%%%%%%%%%%%%%
\begin{figure}[!htb]
  \centerline{\hbox{
      \includegraphics[width=0.7\columnwidth,angle=-90]{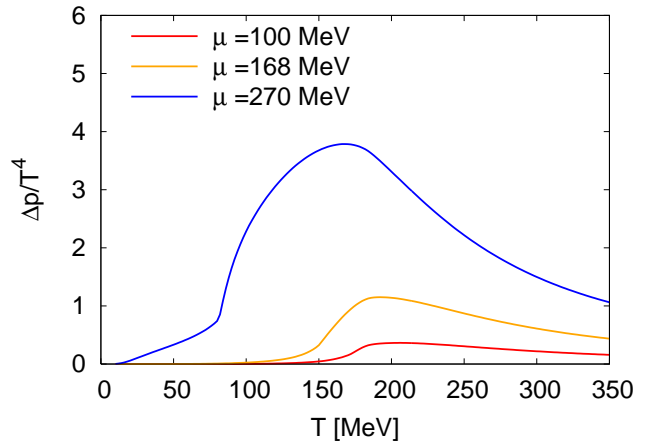}
}}
\caption{\label{fig:pressure_diff} Scaled pressure difference $\Delta
  p/T^4$ for three different chemical potentials. The curves
  correspond to $\mu = 100, 168, 270$ MeV from below.}
\end{figure}
%%%%%%%%%%%%%%%%%%%%%%%%%%%%%%%%%%%%%%%%%%%%%%%%%%%%%%%%%%%
In Fig.~\ref{fig:pressure_diff} the scaled pressure difference $\Delta
p(T,\mu)/T^4$ versus temperature for three chemical potential values
is shown. The bottom curve corresponds to $\mu=100$ MeV. It is always
a continuous function and shows a kink at a first-order phase
transition. $\Delta p$ rises steeply across the chiral transition and
peaks almost at the same temperature for all chemical potentials. For
larger temperatures it decreases almost as $1/T^2$ which follows from
the SB limit. Nevertheless, due to the $T$- and $\mu$-dependent
Polyakov fields slight deviations of the $1/T^2$ SB behavior are seen.

\begin{figure}[!htb]
  \centerline{\hbox{
      \includegraphics[width=0.7\columnwidth,angle=-90]{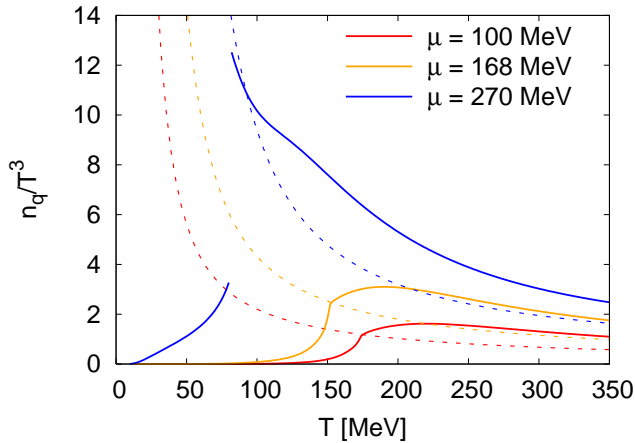}
    }}
  \caption{\label{fig:q_density} Same as described in the legend to
    Fig.~\ref{fig:pressure_diff} for the quark number density
    $n_q/T^3$. The dashed lines denote the corresponding Stefan-Boltzmann
    limits.}
\end{figure}

Another interesting observable is the net quark density. It is
obtained from the thermodynamic potential via $n_q = -
\partial\Omega(T,\mu)/\partial \mu$. The quark density, normalized to
$1/T^3$, is displayed as a function of the temperature in
Fig.~\ref{fig:q_density} for three different chemical potentials
$\mu=100, 168$ and $270$ MeV.
In comparison to the pure quark-meson model without the Polyakov loop
the quark density in the confined phase is much more suppressed when
the interaction of quarks with the Polyakov loop is
added~\cite{Schaefer1999, Schaefer2006a}. A similar effect is seen in
the PNJL model~\cite{Sasaki:2006ww}. Above the phase transition, the
quark density of the pure quark-meson model approaches the
Stefan-Boltzmann limit $n_q = N_f \mu ( T^2 + (\mu/\pi)^2)$
immediately. With the Polyakov loop dynamics this behavior is changed
drastically. The quark densities increase slightly above the
corresponding SB limits and decrease again with growing temperature.
For high temperatures the SB limit of the quark density is always
reached from above. At a first-order phase transition $n_q$ jumps and
drops immediately after the transition for increasing temperatures.

\begin{figure}[!htb]
  \centerline{\hbox{
      \includegraphics[width=0.7\columnwidth,angle=-90]{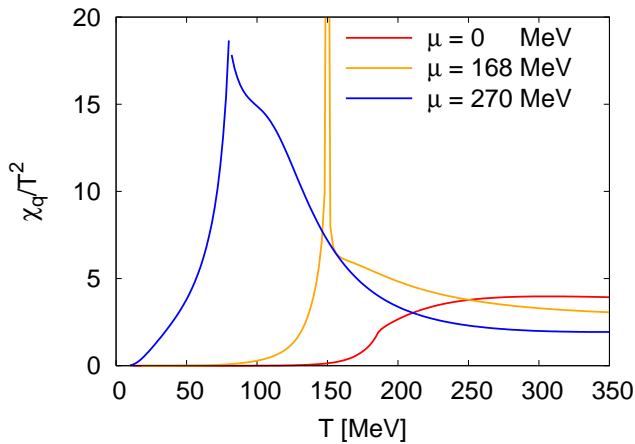}
    }}
  \caption{\label{fig:suscept} The scaled chiral susceptibility
    $\chi_q/T^2$ as a function of temperature for three different
    quark chemical potentials, $\mu =0 , 168, 270$ MeV.}
\end{figure}

The quark number susceptibility measures the static response of the
quark number density to an infinitesimal variation of the quark
chemical potential and is given by $\chi_q = \partial n_q /\partial
\mu$. It is shown in Fig.~\ref{fig:suscept} as a function of
temperature for several $\mu$.
This observable can be used to verify the existence and location of
the critical end point in the phase diagram. At a first-order phase
transition this quantity has a discontinuity and only at a
second-order critical end point it is divergent. Even for finite pion
masses the critical point is of second-order and induces a divergent
quark number susceptibility. This behavior is seen in
Fig.~\ref{fig:suscept}. For $\mu= 168$ MeV, close to the critical
chemical potential, $\chi_q$ diverges at the critical temperature.

The modifications caused by the quark-gluon interaction on the quark
number susceptibility, are similar as those already discussed in the
context of the quark number density. Compared to the pure quark-meson
model $\chi_q$ is again more suppressed below the chiral phase
transition. Above the transition $\chi_q$ lies above the corresponding
SB limit $\chi_q/T^2 = N_f (1 + 3/\pi^2 (\mu/T)^2)$. At high
temperatures the SB limit (not shown in the figure) is again reached
from above.

%%%%%%%%%%%%%%%%%%%%%%%%%%%%%%%%%%%%%%%%%%%%%%%%%%%%%%%%%%%%%%%%%%%%%%%%%%%%%
\section{Conclusion}
\label{sec:conclusion}

In the present paper we have extended the $N_f=2$ quark-meson model to
include certain aspects of gluon dynamics via the Polyakov loop. This
PQM model combines important symmetry aspects in the limit of
infinitely heavy quarks with those of the light quark sector of nearly
massless up- and down quarks. Within the mean-field approximation we
have discussed the ensuing phase diagram of strongly interacting
matter in the grand canonical ensemble, as was done previously in a
similar extension of the NJL model \cite{Roessner:2006xn,
  Ratti:2006wg, Ratti:2007jf}. One of the benefits is an improvement
of the thermodynamical behavior of several bulk quantities such as
pressure, entropy etc. when compared to lattice data. As a novelty we
propose to include the $N_f$ and $\mu$ dependence of the the running
coupling $\alpha$ in the parameter of the Polyakov loop potential. A
qualitative estimate is provided by the one loop $\beta$-function for
the gauge coupling $\alpha$ as well as using the hard dense loop
approximation. Then we are led to a $N_f$ and $\mu$ dependent $T_0$,
the critical temperature of the Polyakov loop model, which decreases
with increasing $N_f$ and $\mu$. These modifications already involve
coinciding peaks in the temperature derivative of the Polyakov loop
expectation value and the chiral condensate at $\mu=0$, in agreement
with the lattice findings of~\cite{Karsch:1994hm, Allton:2002zi}.
Interestingly this coincidence of the deconfinement and chiral
symmetry restoration persists at finite $\mu$.

The findings of the present work provide a promising starting point
for a functional RG study in the present model \cite{Pawlowski2006},
and further extensions towards full QCD: in particular, we aim at
removing the perturbative nature of the above estimates as well as
allowing for a fully coupled PQM model. The last step consists of
including the full gauge dynamics. This is particularly relevant for
the important issue of the Polyakov loop potential at finite $\mu$, 
being intimately related to the open question of the existence and 
location of the critical point in the QCD phase diagram.

\section*{Acknowledgment}

We thank R.~Alkofer, B.~Friman, H.~Gies, R.D.~Pisarski, C.~Ratti,
K.~Redlich, S.~R\"o{\ss}ner, C.~Sasaki, I.O.~Stamatescu, L. von Smekal
and W.~Weise for useful discussions, and the referee for useful comments.

%%%%%%%%%%%%%%%%%%%%%% references %%%%%%%%%%%%%%%%%%%%%%%%%%%%%%%%%%%%%%%

\end{document}